\documentclass{revtex4-2}

\usepackage{amsmath}
\usepackage{graphicx}
\usepackage{multirow}
\usepackage{threeparttable}
\usepackage{tabularx}
\usepackage{float}
\begin{document}

\title{The B[e] Phenomenon in Supergiants. A Result of Mass Transfer in Binaries, Mergers, or What?}

\author{
A.S.~Miroshnichenko$^{1,2,3}$, S.V.~Zharikov$^4$, N.L.~Vaidman$^{2,3}$, S.A.~Khokhlov$^{2,3}$
}

\affiliation{
$^1$ Department of Physics and Astronomy, University of North Carolina Greensboro, Greensboro, NC 27402–6170, USA \\
$^2$ Fesenkov Astrophysical Institute, Observatory 23, Almaty 050023, Kazakhstan \\
$^3$ Al-Farabi Kazakh National University, Faculty of Physics and Technology, 71 Al-Farabi Ave., Almaty 050040, Kazakhstan \\
$^4$ Universidad Nacional Aut\'onoma de M\'exico, Instituto de Astronom\'ia, AP 106, Ensenada 22800, BC, Mexico
}

\begin{abstract}
The B[e] phenomenon discovered nearly 50 years ago features the presence of forbidden emission lines due to extended and dense circumstellar gas and large IR excesses due to the radiation from circumstellar dust in a wide variety of objects from pre-main-sequence stars to Planetary Nebulae. It also shows up in a small group of supergiants that includes Luminous Blue Variables, such as $\eta$ Carinae. Over the years, some of them were proven to be binary systems, but the presence of a secondary component in other is still elusive. At the same time, there is growing evidence that the B[e] phenomenon can be due to binary mergers or interactions in triple systems.
\end{abstract}


\maketitle


\section{Introduction}

There are phases in stellar evolution when large amounts of circumstellar (CS) matter are found near stars of all masses. Such phases may last a different amount of time, depending on the star's mass or the arrangement of components in binary/multiple systems. The CS material may be represented by neutral or ionized gas and ensembles of molecules combined in dusty particles. In massive stars, possible formation mechanisms of the CS matter include stellar winds, and mass transfer between components in multiple systems, as the pre-main-sequence stage of their evolution is typically hidden by the natal star-forming cloud.

Although there are many early-type supergiants in the Milky Way, only a small fraction of them exhibit the B[e] phenomenon.
The phenomenon was discovered almost 50 years ago by \citet{1976AA....47..293A} in 65 Galactic Be stars and planetary nebulae and defined as the presence of forbidden (e.g., [O {\sc i}], [Fe {\sc ii}], [N {\sc ii}]) and permitted emission lines (e.g., H {\sc i}, Fe {\sc ii}) in the optical spectra of B--type stars along with large IR excesses due to CS dust.

Initial studies revealed that these properties were found in some members of five stellar groups: pre-main-sequence Herbig Ae/Be stars, symbiotic binaries, compact planetary nebulae, supergiants, and FS\,CMa type objects \citep{1998A&A...340..117L,2007ApJ...667..497M}. The nature and evolutionary state of the former four groups were already known, while the fifth group was recognized nearly two decades ago on the basis of common features (strong line emission and a fast decline of the IR flux past $\lambda \sim 10-30~\mu$m toward longer wavelengths). These features were hypothesized to be due to the CS matter created in ongoing or past mass-transfer in binary systems, because single stars with the observed luminosities from a few hundred to $\sim$30,000 L$_{\odot}$ that correspond to a mass range ($\sim2-20$ M$_{\odot}$) seem to be incapable of producing the sufficient amount of CS material \citep[see][for a recent review]{2023Galax..11...36M}.

Supergiants with the B[e] phenomenon (hereafter sgB[e]) are defined by locations in the Hertzsprung-Russell diagram (HRD) beyond the main-sequence and high luminosities \citep[$\log$ L/L$_{\odot} =5.1\pm0.2$,][]{2007ApJ...667..497M}.
Their fraction in the original list of Galactic objects \citep{1976AA....47..293A} was small. \citet{1998A&A...340..117L} mentioned 2 bona-fide such objects (CPD-52$^{\circ}$9243 and GG\,Car) and 3 probable ones (MWC\,300, MWC\,349A, and HD\,87643). \citet{2006ASPC..355...13M} listed 6 objects ($\eta$\,Car, MWC\,349, CPD-52$^{\circ}$9243, MWC\,300, RY\,Sct, and 3\,Pup), while mentioning seven (probably GG\,Car was mistakenly forgotten). 

The differences in classifying objects as sgB[e]s were mostly due to uncertainties in their luminosities that stemmed from unknown distances and veiling of photospheric lines by a strong CS continuum. It has been much easier to classify similar objects in the Magellanic Clouds because of a well-known distance and a low interstellar extinction. Nearly a dozen of sgB[e]s were discovered in the LMC and SMC by \citet{1986A&A...163..119Z}. The locations of LMC and SMC sgB[e]s in the HRD are shown in Figure~1 of \citep{1998A&A...340..117L}. The Galactic sgB[e]s were typically referred to as analogs of those from the Clouds. A recent survey of sgB[e]s found beyond the Milky Way was published by \citet{2019Galax...7...83K}.

$\eta$\,Car was historically considered a sgB[e] before it was classified as a Luminous Blue Variable (LBV). Some other Galactic LBVs (e.g., AG\,Car and HR\,Car) also exhibit IR excesses, but only at $\lambda \sim 60-100~\mu$m. The absence of a near-IR excess resulted in not classifying them as objects with the B[e] phenomenon.


\section{Enlarging the original list of sgB[e]}

Although the number of stars in the original search done by \citet{1976AA....47..293A} was large (700 objects), some potential B[e] candidates went missing. In particular, an object that was later recognized to show the B[e] phenomenon (HD\,327083) was found by \citet{1979VA.....23..213C}, a few more were revealed by \citet{1991AcApS..11..172D} using a positional cross-correlation of a catalog of Galactic early-type emission-line stars \citep{1970MmRAS..73..153W} and the IRAS Point Source Catalog \citep{1988iras....7.....H}. For a more detailed description of these findings, see \citet{2007ApJ...667..497M}.

The IRAS survey expanded the wavelength coverage of the stellar spectral energy distributions further in the IR spectral region that allowed finding new observational features of objects with the B[e] phenomenon. \citet{2007ApJ...667..497M} analyzed IRAS fluxes of these objects from \citet{1976AA....47..293A} and \citet{1991AcApS..11..172D} and found that a noticeable number of them exhibited a strong decrease in flux at $\lambda = 60~\mu$m \citep[see Figure~1 in][]{2007ApJ...667..497M}. This feature pointed to a lack of cold dust around the stars and provided a separation criterion from pre-main-sequence objects, which still retained distant remnants of their protostellar clouds, and part of the foundation of the FS\,CMa group (see above). This study along with several previous investigations of individual objects conducted in the early 2000's (see Table\,\ref{tab1}) resulted in enlarging the list of Galactic sgB[e].

At that time, determination of the objects' luminosities was primarily based on spectroscopic distances, as the parallaxes measured by the HIPPARCOS mission were either too uncertain or unavailable due to large distances and low brightnesses of the majority of objects with the B[e] phenomenon. Nevertheless, diffuse interstellar bands present in the objects' spectra and existing relationships of the interstellar extinction with distance in their directions allowed reasonably reliable distance and hence luminosity estimates. Currently, Gaia DR3 parallaxes \citep{2021AJ....161..147B} provided more accurate distances for some sgB[e] that still need to be checked for systematic errors due to the presence of possible stellar companions or asymmetric CS structures. The existing data for the most recognized Galactic sgB[e]s are summarized in in Table\,\ref{tab1}.

\begin{table*}[ht]
\centering
\caption{Parameters of a sample of Galactic sgB[e] objects}
\label{tab1}

\begin{ruledtabular}
\begin{tabular}{lccccccc}
\multicolumn{1}{c}{1} &
\multicolumn{1}{c}{2} &
\multicolumn{1}{c}{3} &
\multicolumn{1}{c}{4} &
\multicolumn{1}{c}{5} &
\multicolumn{1}{c}{6} &
\multicolumn{1}{c}{7} &
\multicolumn{1}{c}{8} \\
\hline
Object ID & $V$ (mag) & $\log T_{\rm eff}$ (K) & $D_{\rm sp}$ (kpc) & $D_{\rm Gaia}$ (kpc) 
& $E(B-V)$ (mag) & $\log L/L_\odot$ & Ref \\
\hline
Hen\,3--298        & 12.6--12.7 & 4.30 & 4.0 & 6.64$^{+1.23}_{-0.86}$ & 1.7 & 5.1 / 5.38$\pm0.16$ & 1 \\
Hen\,3--303        & 14.1--14.3 & 4.30 & 4.0 & 7.76$^{+0.79}_{-0.94}$ & 1.7 & 4.3 / 4.85$\pm0.14$ & 1 \\
HD\,327083         & 9.7--10.0  & 4.30 & 1.5 & 2.26$^{+0.14}_{-0.11}$ & 1.8 & 5.0 / 5.68$\pm0.11$ & 2 \\
Hen\,3--1398       & 10.5--10.8 & 4.47 & 3.3 & 1.80$^{+0.07}_{-0.10}$ & 1.1 & 5.3 / 4.69$\pm0.10$ & 1 \\
MWC\,300           & 11.5--11.7 & 4.25 & 1.8 & 1.68$^{+0.07}_{-0.08}$ & 1.2 & 5.1 / 3.87$\pm0.08$ & 3 \\
AS\,381            & 14.0--14.6 & 4.28 & 4.0 & 6.20$^{+1.40}_{-0.45}$ & 2.2 & 4.9 / 5.24$\pm0.18$ & 1 \\
RY\,Sct            & 9.0--9.7   & 4.43 & 1.8 & 1.96$^{+0.09}_{-0.08}$ & 1.3 & 5.6 / 5.41$\pm0.16$ & 4 \\
GG\,Car            & 8.4--8.9   & 4.36 & 3.4 & 2.41$^{+0.71}_{-0.92}$ & 0.5 & 5.3 / 4.75$\pm0.15$ & 5 \\
HD\,87643          & 8.9--9.8   & 4.26 & 1.2 & 1.59$^{+0.29}_{-0.31}$ & 0.6 & 3.3 / 3.97$\pm0.41$ & 6 \\
3\,Pup             & 3.9--4.0   & 3.93 & 0.7 & 1.02$^{+0.23}_{-0.12}$ & 0.1 & 4.1 / 4.40$\pm0.20$ & 7 \\
MWC\,349A          & 13.1--13.2 & 4.40 & 1.7 & 1.46$^{+0.23}_{-0.12}$ & 2.8 & 5.7 / 5.46$\pm0.15$ & 8 \\
CPD$-52^{\circ}9243$  & 10.1--10.5 & 4.20 & 3.4 & 3.88$^{+0.23}_{-0.27}$ & 1.7 & 5.0 / 5.63$\pm0.15$ & 9 \\
CPD$-57^{\circ}2874$  & 10.0--10.4 & 4.30 & 2.5 & 4.50$^{+0.39}_{-0.33}$ & 1.8 & 5.7 / 6.16$\pm0.15$ & 10 \\
MWC\,137           & 12.0--12.1 & 4.48 & 6.0 & 4.78$^{+0.69}_{-0.66}$ & 1.5 & 5.4 / 5.49$\pm0.15$ & 11 \\
\end{tabular}
\end{ruledtabular}

Column information: 1 -- an example of the object's ID, 2 -- $V$--band brightness limits from the ASAS SN database \citep{2017PASP..129j4502K}, 3 -- effective temperature in K, 4 -- spectroscopic distance in kpc, 5 -- Gaia DR3 distance in kpc, 6 -- color-excess from combined CS and interstellar extinction, 7 -- bolometric luminosity derived from spectroscopy versus one from the Gaia DR3 distance, 8 -- Reference to the fundamental parameters derived from spectroscopy:\\ 1 -- \cite{2007ApJ...667..497M}, 2 -- \cite{2024ApJ...968...52N}, 3 -- \cite{2012A&A...545L..10W}, 4 -- \cite{2005A&A...443..211M}, 5 -- \cite{2021MNRAS.501.5554P}, 6 -- \cite{2009A&A...507..317M}, 7 -- \cite{2020ApJ...897...48M}, 8 -- \cite{2002A&A...395..891H}, 9 -- \cite{2012A&A...548A..72C}, 10 -- \cite{2007A&A...464...81D}, 11 -- \cite{1998MNRAS.298..185E}.\\
Note: The luminosity derived from the Gaia DR3 distance also uses the ASAS SN average $V$--band magnitude and bolometric corrections appropriate for the T$_{\rm eff}$ shown in column 3.
\end{table*}

\section{Recognized binary systems and objects with uncertain classification}

Most objects listed in Table\,\ref{tab1} were found or suspected to be binary systems. Orbital periods have been derived for HD\,327083 \citep[107.7 days,][]{2024ApJ...968...52N}, RY\,Sct \citep[11.1 days,][]{2001A&A...374..638D}, GG\,Car \citep[31.01 days,][]{2021MNRAS.501.5554P}, 3\,Pup \citep[137.5 days,][]{2025Galax..13..101V}. Signatures of a K-type companion were found in the spectrum of AS\,381 \citep{2002A&A...383..171M}, \citet{2009A&A...507..317M} detected a companion of HD\,87643 with interferometry, and \citet{2012A&A...545L..10W} suspected binarity in MWC\,300. 

Another example of a binary system not included in Table\,\ref{tab1} is CI\,Cam from the original list of B[e] stars. Its orbital period of 19.41 days \citep{2006ARep...50..664B} was suggested from the detection of periodic positional variations of the He {\sc ii} 4686 \AA\, emission line, which is supposed to form in the CS medium of the secondary companion that is much fainter than the $V \sim 11.6$ mag primary. This object has been considered a sgB[e] in many studies, although the distance toward it was estimated from 1 to $\sim$10 kpc. Recent analysis of TESS data showed that part of the photometric variability is due to radial pulsations that constrained the primary component's mass and ruled out its supergiant status \citep{2023AstBu..78....1B}.

At the same time, it is still unclear whether MWC\,349A, a close visual binary (separation 2.4 arcsec), is physically connected to a fainter companion MWC\,349B \citep{2017ASPC..508..389M}. The remaining objects (Hen\,3--298, Hen\,3--303, CPD$-52^{\circ}9243$, and CPD$-57^{\circ}2874$) show no convincing signs of binarity.

Some of the mentioned objects do not seem to be genuine massive stars, because of potential effects of mass transfer in binaries that may significantly alter their masses and luminosities. For example, 3\,Pup shows an absorption-line spectrum of an A3--type supergiant, but its evolutionary history suggests that this binary began its evolution as a 6.0+3.6 M$_{\odot}$ pair, where the initially less massive companion gained nearly 5 M$_{\odot}$ from the more massive one, which became an extremely hot (T$_{\rm eff} \sim 50,000$ K) compact star with a less than 1 M$_{\odot}$ mass \citep{2020ApJ...897...48M}. A much larger Gaia DR3 distance compared to the spectroscopic one \citep[although uncertain, see][]{2005A&A...436..653M} brought the luminosity of Hen\,3--303 very close to the average one for sgB[e]s. At the same time, although neither spectroscopic nor Gaia luminosity of HD\,87643 is close to those of sgB[e]s, it is still considered a member of the sgB[e] group \citep{2009A&A...507..317M}.

The evolutionary stage of a few objects is still uncertain. MWC\,137, MWC\,300, Hen\,3--303, and HD\,87643 still appear in studies of pre-main-sequence stars \citep[e.g.,][]{2023A&A...679A..71S}. However, MWC\,137, which is surrounded by a large nebula Sh\,2--266, was rather convincingly shown to be a high-luminosity object by \citet{1998MNRAS.298..185E} and \citet{2015AJ....149...13M}. Also, Hen\,3--303 is listed in the SIMBAD database as a Long-Period Variable candidate. All these facts reflect complicated properties of the Galactic population of sgB[e]s.

\section{Nature of Galactic sgB[e]s}

Numerous studies of Galactic B-type stars of high luminosity without CS dust \citep[e.g.,][to name a few]{1999A&AS..137..351V,2022A&A...668A..92W} show that those with $\log$ L/L$_{\odot} \ge 5.0$ exhibit emission-line spectra that are typically represented by the H$\alpha$ line only. Since the B[e] phenomenon is evident in only a very small fraction of such stars, certain special conditions need to be present in these systems. One possibility is binarity, which has been proven to be the case in a big fraction of the sgB[e] group members and candidates. More specifically, only binary systems of a certain range of masses and orbital separations may pass through this evolutionary stage.

Another possibility that has recently been suggested for explaining the properties of objects with the B[e] phenomenon invokes scenarios with merging stars. \citet{2021MNRAS.503.4276H} proposed to explain the formation of the Homunculus nebula around $\eta$ Carinae by a stellar merger in an unstable triple system. \citet{2023MNRAS.523.5554M} investigated MHD simulations to explain disk-like CS envelopes of FS\,CMa objects, which these authors consider to be post-mergers in binary systems. A number of sgB[e] described in this paper is mentioned as potential merger products by the latter authors.

\section{Conclusions}

The B[e] phenomenon is a remarkable manifestation of a hot star radiation processed by large amounts of CS matter distributed in an extended area. Investigation of all groups that exhibit the phenomenon expands our knowledge of stellar evolution and triggers exciting discoveries. Despite recent progress in studying stars with various methods, even those discovered long ago continue to present new puzzles and testing our understanding of physical processes that take place in stars and their environments.
The complexity and dynamics of many B[e] objects require frequent attention with a multitude of observing techniques and sophisticated modeling tools.

In particular, frequent all-sky photometric surveys provided constraints on the optical brightness variability of sgB[e]s. However, a lack of medium- and high-resolution spectroscopic observations hampers studies of spectral line positional variations and detections of possible companions, while rare photometric observations in the IR spectral region prevent studies of the temporal behavior of CS matter. Although most bright sgB[e]s ($V \le 12$ mag) have been found, there are still potentially undiscovered such objects among fainter stars with IR excesses. Another outcome of recent studies of B[e] objects is that both types of them, which produce CS dust during their recent or current evolutionary stages (sgB[e]s and FS\,CMa objects), may represent products of evolution of binary or multiple systems with a certain range of masses and orbital separations.

\section{Funding}

This research was funded by the Science Committee of the Ministry of Science and Higher Education of the Republic of Kazakhstan (Grant No. AP23484898).
A.~S.~M. acknowledges support from the College of Arts and Sciences and the Department of Physics and Astronomy of the University of North Carolina Greensboro.

\end{document}